\documentclass[article]{jss}

\usepackage{thumbpdf,lmodern}

\usepackage{framed}

\usepackage{amsmath, bm, dsfont}

\newcommand{\fct}[1]{\code{#1()}}
\newcommand{\bx}{{\bm x}}
\newcommand{\by}{{\bm y}}

\newcommand{\bbeta}{{\bm \beta}}
\newcommand{\bmu}{{\bm \mu}}
\newcommand{\bpsi}{{\bm \psi}}
\newcommand{\bxi}{{\bm \xi}}
\newcommand{\bpsiu}{{{\bm \psi}_u}}
\newcommand{\bu}{{\bm u}}
\newcommand{\bsig}{{\bm \sigma}}
\newcommand{\acite}[1]{\citeauthor{#1}' (\citeyear{#1})}
\newcommand{\ascite}[1]{\citeauthor{#1}'s (\citeyear{#1})}

\author{Benjamin D. Youngman\\University of Exeter}
\Plainauthor{Benjamin D. Youngman}

\title{\pkg{evgam}: An \proglang{R} package for Generalized Additive Extreme Value Models}
\Plaintitle{evgam: Generalized Additive Extreme Value Models in R}
\Shorttitle{\pkg{evgam}: Generalized Additive Extreme Value Models in \proglang{R}}

\Abstract{
This article introduces the \proglang{R} package \pkg{evgam}. The package provides functions for fitting extreme value distributions. These include the generalized extreme value and generalized Pareto distributions. The former can also be fitted through a point process representation. \pkg{evgam} supports quantile regression via the asymmetric Laplace distribution, which can be useful for estimating high thresholds, sometimes used to discriminate between extreme and non-extreme values. The main addition of \pkg{evgam} is to let extreme value distribution parameters have generalized additive model forms, the smoothness of which can be objectively estimated using Laplace's method. Illustrative examples fitting various distributions with various specifications are given. These include daily precipitation accumulations for part of Colorado, US, used to illustrate spatial models, and daily maximum temperatures for Fort Collins, Colorado, US, used to illustrate temporal models.
}

\Plainkeywords{Generalized extreme value distribution, generalized Pareto distribution, point process, generalized additive model, Laplace's method, R}

\Address{
  Benjamin D. Youngman\\
  Department of Mathematics\\
  University of Exeter\\
  Laver Building, North Park Road\\
  Exeter, EX4 4QE, UK\\
  E-mail: \email{b.youngman@exeter.ac.uk}\\
  URL: \url{http://ex.ac.uk/youngman}
}

\begin{document}




\section[Introduction]{Introduction} \label{sec:intro}

Practical extreme value analyses have typically considered modeling block maxima with the generalized extreme value (GEV) distribution or exceedances of a high threshold using the generalized Pareto distribution (GPD); see \cite{davison1990} for a seminal work on the latter approach, and \cite{coles2001} for a detailed overview of both approaches. Here, the GEV and GPD distributions will be considered \emph{the} extreme value distributions (EVD). \cite{smith1989} develops a model using \acite{pickands1971} point process representation of extremes, which, in some sense, marries the two EVDs.

Various packages have been contributed to the Comprehensive \proglang{R} Archive Network (CRAN) to fit EVDs in \proglang{R} \citep{R}. One of the earliest, \pkg{ismev} \citep{ismev}, allows users to recreate many of the analyses presented in \cite{coles2001}. Later \proglang{R} packages, such as \pkg{evd} \citep{evd}, \pkg{evir} \citep{evir}, \pkg{extRemes} \citep{gilleland2016} and \pkg{mev} \citep{mev}, have offered various functions for fitting univariate and multivariate EVDs. For a review see \cite{gilleland2013}, and for an up-to-date list of packages contributed to CRAN see \ascite{dutang2020} Task View.

This work focuses on regression-based models for extremes, a flexible class of nonstationary model for extremes achieved by letting EVD parameters vary with covariates. Nonstationarity was considered in early models for extremes, in particular \cite{smith1986} and \ascite{smith1989} study of trends ground-level ozone. Packages \pkg{ismev} and \pkg{evd} offer some scope for linear forms. Such forms, however, can be restrictive if involved choice of covariate parametrization is required before sufficient flexibility can be achieved (if it can).

More general regression-based EVD parameter forms can offer more robust analyses. \cite{hall2000}, \cite{davison2000} and \cite{ramesh2002}, for example, considered local-likelihood fitting of trends. \cite{pauli2001} used a penalized likelihood in which smoother EVD parameter estimates incur less penalty. \ascite{pauli2001} approach builds on results for exponential family models covered in \cite{green1994}, but relies on fixed smoothing parameters to control the amount of penalty. \cite{chavez2005} consider generalized additive model (GAM) forms for GPD parameters, which allow a given parameter to be represented with one or more `smooths', i.e., smooth functions, each of which may have a different smoothness. \cite{yee2007} consider the vector GAM (VGAM) setting of \cite{yee1996} for representing EVD parameters with GAM form. More recently, \cite{randell2016} use spline forms and roughness-penalized priors to represent variation in EVD parameters when modeling significant wave heights, using Markov chain Monte Carlo for inference. \cite{y2019} models exceedances of a threshold with a GPD with parameters of GAM form and a high threshold estimated by GAM form quantile regression, as proposed in \cite{yee2007} and \cite{northrop2011}.

GAM forms typically consider additive smooths represented with splines. Various packages contributed to \proglang{R} fit EVDs with parameters of GAM or spline form. In particular, \pkg{VGAM} allows the GEV and GPD distributions to be fitted with parameters of GAM form. Various EVDs are also available within \pkg{gamlss} \citep{rigby2005}. Alternatively, \pkg{ismev}'s \fct{gamGPDfit} implements \cite{chavez2005}, i.e. fits a GPD with parameters of GAM form through backfitting. Marginal spline forms are also allowed for GEV parameters in \pkg{SpatialExtremes} \citep{SpatialExtremes}, although the package's focus is multivariate analyses, in particular with max-stable processes. Fitting of the GEV with parameters of GAM forms is also possible with \pkg{mgcv} with option \code{family = `gevlss'}. EVDs can also be fitted using the integrated nested Laplace approximation (INLA) software \cite{rue2009}, which specifies smooths as latent Gaussian random fields (GRF) that depend on hyperparameters. Options for GAM-based quantile regression, which can be useful for threshold estimation, include \pkg{VGAM} and \pkg{qgam} \citep{fasiolo2020}. \pkg{quantreg} \citep{quantreg} allows quantile regression using B-splines.

Estimating GAM forms for EVDs under fixed smoothing penalties is fairly straightforward. For example, parameter estimates can maximize a penalized log-likelikood; recall \cite{pauli2001}. Smoothing parameter (or hyperparameter) selection, however, is perhaps the most challenging part of fitting a distribution with parameters of GAM form. In \pkg{VGAM} this is eased by users specifying the degrees of freedom for smooths, from which smoothing parameter estimates are derived. Degrees of freedom are deemed more intuitive to specify than smoothing parameters themselves. Alternatively \pkg{gamlss} includes \fct{find.hyper}, which minimizes a generalized Akaike information criterion to find optimal degrees of freedom. \cite{wood2011} proposes an objective approach to smoothing parameter estimation for exponential family distributions by treating penalized parameters as multivariate Gaussian random effects, which are integrated out by Laplace's method. This gives a marginal likelihood for smoothing parameters; see \S\ref{reml}. \cite{wood2016} extend this approach beyond the exponential family. This method is implemented in \fct{gam} from \pkg{mgcv} with option \code{method} set to \code{"ML"} or \code{"REML"}. Laplace's method is used by \cite{rue2009} in INLA to integrate out latent GRFs so that hyperparameters can be optimally estimated. Optimal estimation can be beneficial when degrees of freedom cannot easily be user-specified: a GAM form comprising many smooths is one example. 

The aim of \pkg{evgam} is to bring together three things: 1) the flexibility of the different smooths available in \pkg{mgcv} for fitting EVDs with parameters of GAM form; 2) objective inference for \emph{all} parameters; and 3) functions for drawing common inferences from extreme value analyses, such as return level estimates with uncertainty quantified. For 1), in particular, \pkg{mgcv} offers GAMs incorporating thin plate regression splines, which are particularly attractive for modeling multidimensional processes, such as spatial processes, or interactions between splines formed by tensor products, as implemented in \pkg{mgcv} through \fct{te}. For 3), \pkg{evgam} provides functionality for estimating return levels from nonstationary EVD parameters and straightforward quantification of their uncertainty. 

Initially \pkg{evgam} performed the analysis of \cite{y2019}, i.e., using the asymmetric Laplace distribution (ALD) to estimate a quantile of GAM form, and then estimating the distribution of its excesses with a GPD with parameters of GAM form. This article presents extensions that allow estimation of GEV distribution parameters of GAM form. These can be estimated from block maxima or from threshold exceedances through the point process model of \cite{smith1989}. The point process model allows simultaneous estimation of all parameters required for return level estimation, while potentially being less wasteful of data than the block maxima approach. Furthermore, the point process model is implemented through the intuitive $r$-largest order statistics model representation; see, e.g., \citet[\S7.9]{coles2001}. Finally, \pkg{evgam} allows estimation of EVDs based on censored data, which can be useful for data known to be recorded with little precision, and is also available in \pkg{gamlss}.

The next section gives details of EVDs available in \pkg{evgam}, deriving return levels from them, and a summary of how they are fitted. Section \ref{sec:evgam} introduces \pkg{evgam}'s main functions. Section \ref{sec:illustrations} presents various examples of use of \pkg{evgam}, including spatial and temporal models. A brief summary is given in Section \ref{sec:summary}.

\section{Extreme value modeling} \label{sec:models}

\subsection{Extreme value distributions} \label{sec:models:evd}

This section outlines the three EVD models supported by \pkg{evgam}, and quantile regression via the ALD; see \cite{yu2001}. Fuller details of the EVD models can be found in \citet[chapters 3, 4 and 7]{coles2001}.

\subsubsection{Generalized extreme value distribution} \label{sec:models:gev}

The GEV distribution is appropriate for block maxima of sufficiently large blocks. Here years will be considered as blocks, to help intuition; henceforth we will refer to \emph{annual} maxima. A random variable $Y$ that is GEV distributed has cumulative distribution function (CDF) 
\[ 
F_\text{GEV}(y; \mu, \psi, \xi) = \exp\left\{-\left[1 + \xi\left(\dfrac{y - \mu}{\psi}\right)\right]^{-1/\xi}\right\}, 
\] 
which is defined for $\{y : 1 + \xi (y - \mu) / \psi > 0\}$ with $(\mu, \psi, \xi) \in \mathds{R} \times \mathds{R}^+ \times \mathds{R}/\{0\}$. The limit $\xi \to 0$ is used for the $\xi = 0$ case, which corresponds to the Gumbel CDF, $\exp(-\exp\{-[(y - \mu)/\psi]\})$. For all models this limit is invoked in \pkg{evgam} if $|\xi| < 10^{-6}$.

\subsubsection{Generalized Pareto distribution} \label{sec:models:gpd}

The GPD is used to model excesses of a high threshold $u$. For a random variable $Y$, it is a model for the conditional distribution $(Y-u) \mid (Y > u)$ with CDF
\[
F_\text{GPD}^{(u)}(y; \psi_u, \xi) = 1 - \left[1 + \xi\left(\dfrac{y}{\psi_u}\right)\right]^{-1/\xi},
\]
which is defined for $\{y : y > 0 \text{ and } 1 + \xi y / \psi_u > 0\}$ with $(\psi_u, \xi) \in \mathds{R}^+ \times \mathds{R}/\{0\}$. The exponential CDF, $1 - \exp(-y/\psi_u)$, is used for the $\xi = 0$ case.

\subsubsection{Poisson-GPD point process model} \label{sec:models:pp}

The Poisson-GPD point process model is considered as an extension of the GPD model with high threshold $u$ that allows estimation of GEV parameters. For random variables $\{Y_i\}_{i = 1, \ldots, n}$ and $y > u$ the Poisson-GPD model has intensity measure
\[
\Lambda(A) = n_y (t_2 - t_1) \left[1 + \xi\left(\dfrac{y - \mu}{\psi}\right)\right]^{-1/\xi}
\]
where $A = [t_1, t_2] \times (y, \infty)$, $n_y$ is the time period under study and $t_i = (i - 0.5) / n$.

\subsubsection{Asymmetric Laplace distribution (for threshold estimation)} \label{sec:models:ald}

The ALD is not an EVD in the usual sense. It is useful in threshold-based extreme value analyses for allowing quantile estimation \citep{yu2001}. The GPD and Poisson-GPD models rely on a `high' threshold. \citet[Chapter 4]{coles2001} discusses assessing its choice. High can be sometimes be intuitively defined through a high quantile, e.g., 0.9, 0.95 or 0.99. Quantile regression can be used to estimate such thresholds, especially covariate-dependent thresholds. The ALD has density function
\[
f_{\text{ALD}, \tau}(y; u, \sigma, \tau) = \dfrac{\tau(1 - \tau)}{\sigma} \exp\left\{-\rho_\tau\left(\dfrac{y - u}{\sigma}\right)\right\},
\]
where $\rho_\tau(y) = y(\tau - I\{y < 0\})$ is the check function, for indicator function $I\{\}$; see \cite{koenker2005} for an overview of quantile regression. The modified check function of \cite{oh2011} is used in \pkg{evgam} to ease inference.

\subsection{Return levels}

Return levels are often sought from extreme value analyses. If the the annual maximum has CDF $F_\text{ann}$, say, then the return level, $z_p$, corresponding to return period $1/p$ years, satisfies $F_\text{ann}(z_p) = 1 - p$.

\subsubsection{GEV and Poisson-GPD models}

For the GEV distribution
\begin{equation} \label{gev_rl}
z_p = \mu - \frac{\psi}{\xi} \left\{1 - \left[-\log(1 - p)\right]^{-\xi}\right\},
\end{equation}
when $\xi \neq 0$ and $\mu - \psi \log(-\log(1 - p))$ otherwise. Eq. \eqref{gev_rl} also applies to the Poisson-GPD model if formulated in terms of annual maxima.

\subsubsection{GPD model}

For a GPD representing independent excesses of threshold $u$, where $n_y$ observations occur each year and such that $\text{Pr}(Y > u) = \zeta$,
\begin{equation} \label{gpdrl}
z_p = u + \frac{\psi_u}{\xi} [(n_y \zeta / p)^{\xi} - 1],
\end{equation}
when $\xi \neq 0$ and $u + \psi_u \log(n_y \zeta / p)$ otherwise.

For the GEV it is typically reasonable to assume that annual maxima are independent. For the GPD, however, excesses of a threshold may occur in clusters, which requires that Eq. \eqref{gpdrl} be adjusted accordingly. This is achieved through the extremal index, $0 < \theta \leq 1$, so that $z_p = u + \frac{\psi_u}{\xi} [(n_y \zeta \theta / p)^{\xi} - 1]$ when $\xi \neq 0$ and $u + \psi_u \log(n_y \zeta \theta / p)$ otherwise. Currently \pkg{evgam} only allows relatively simple, constant estimates of $\theta$ based on the moment-based estimator of \cite{ferro2003}. An example is given in \S\ref{FC:dy}.

\subsection{Nonstationarity}

\subsubsection{Outline}

\sloppy Now consider $Y(\bx)$, a random variable indexed by covariate $\bx$. The purpose of \pkg{evgam} is to allow straightforward fitting of EVDs with parameters that vary flexibly with $\bx$. The following notation will be used. For the GEV, suppose that annual maxima $Y(\bx) \sim GEV\big(\mu(\bx), \psi(\bx), \xi(\bx)\big)$; for the GPD, that $Y(\bx) - u(\bx) \mid Y(\bx) > u(\bx) \sim GPD\big(\psi_u(\bx), \xi(\bx)\big)$; for the Poisson-GPD model, that $Y(\bx) - u(\bx) \mid Y(\bx) > u(\bx)$ will be used to estimate $GEV\big(\mu(\bx), \psi(\bx), \xi(\bx)\big)$; and for the ALD that $Y(\bx) \sim ALD\big(u(\bx), \sigma(\bx)\big)$. 

\subsubsection{Return levels} \label{sec:models:rlns}

If covariate $\bx$ relates to time, return levels typically need different treatment. Two examples are given here for illustration: one for the GEV case, and one for the GPD case. These should be sufficient to inform other situations.

Suppose that covariate $\bx$ defines month, i.e, $\bx_i = \text{month}(i)$, for $i=1, \ldots, n$, and that $Y(\bx_i) \sim GEV\big(\mu(\bx_i), \psi(\bx_i), \xi(\bx_i)\big)$ are monthly maxima, which may have a different distribution each month. The CDF of the annual maximum then takes the composite form
\begin{equation} \label{Fann_gev} 
F_\text{ann}(z) = \prod_{\bx_j=1}^{n_y} \big\{F_\text{GEV}\big(z; \mu(\bx_j), \psi(\bx_j), \xi(\bx_j)\big)\big\}^{n_y w(\bx_j)},
\end{equation}
where $n_y=12$ and $w(\bx_j)$ are weights: $w(1) = w(3) = w(5) = w(7) = w(8) = w(10) = w(12) = 31 / 365$, $w(2) = 28 / 365$ and $w(4) = w(6) = w(9) = w(11) = 30 / 365$. (This, for simplicity, considers only 365-day years.) The $1/p$-year return level, $z_p$, satisfies $F_\text{ann}(z_p) = 1 - p$. Unless $z_p$ has closed form, which is rare, it must be found numerically. This approach to return level estimation is implemented in \S\ref{FC:mn}.

The case of covariate $\bx$ being time-dependent is handled similarly with the GPD. Now suppose $\bx_i = \text{day}(i)$. The composite form for $F_\text{ann}$ is then given by
\[
F_\text{ann}(z) = \prod_{\bx_j=1}^{n_y} \big\{F_\text{GPD}\big(z; \zeta(\bx_j), \psi_u(\bx_j), \xi(\bx_j)\big)\big\}^{n_yw(\bx_j)},
\]
where $n_y = 365$ and $F_\text{GPD}$ denotes the unconditional distribution of a random variable $Y$: \begin{equation} \label{Fann_gpd} F_\text{GPD}(y; \zeta, \psi_u, \xi) = 1 - \zeta\left[1 - F_\text{GPD}^{(u)}(y - u; \psi_u, \xi)\right],~~~~\text{for}~y > u,\end{equation} and $\zeta = \text{Pr} (Y > u)$. Here we would take $w(\bx_j)=1 / n_y$, for all $\bx_j$. Again $F_\text{ann}(z_p) = 1-p$ for $z_p$ is typically only solved numerically. This approach is demonstrated in \S\ref{FC:dy} and, additionally, continuous time-dependent $\bx$ is considered. Then infinitely many values exist for $\bx$. $F_\text{ann}$ formed over a product would therefore be an approximation based on the 365-point set $\{1, \ldots, 365\}$. More or fewer points may benefit this approximation's accuracy and computational cost. A 50-point set is used in \S\ref{FC:dy}. 

In the above, composite forms for $F_\text{ann}$ are easily modified for non-monthly maxima or non-daily threshold exceedances. For example, the former might instead use `seasonal' maxima, where season may be problem-specific, or the latter might use hourly threshold exceedances. The return period need not be defined in terms of years, either.

\subsection{Inference} \label{lik}

For the GEV model, consider annual maxima $\{Y(\bx_i)\}_{i = 1, \ldots, n}$. We might obtain these by dividing a sequence of random variables by year and retaining each year's maximum. Let $f_*$ denote a model's density function. The GEV likelihood is then
\[
L(\bmu, \bpsi, \bxi) = \prod_{i=1}^{n} f_\text{GEV}\big(y(\bx_i)); \mu(\bx_i), \psi(\bx_i), \xi(\bx_i)\big),
\]
with $\bmu = (\mu(\bx_1), \ldots, \mu(\bx_n))$, $\bpsi = (\psi(\bx_1), \ldots, \psi(\bx_n))$ and $\bxi = (\xi(\bx_1), \ldots, \xi(\bx_n))$. For the GPD, now let $\{Y(\bx_i)\}_{i=1, \ldots, n}$ be $n$ threshold excesses. The would be obtained by retaining the threshold exceedances from a sequence of random variables and then calculating their excesses of the threshold. The GPD model likelihood is \[ L(\bpsiu, \bxi) \prod_{i=1}^{n} f_\text{GPD}\big(y(\bx_i); \psi_u(\bx_i), \xi(\bx_i)\big), \] with $\bpsi_u = (\psi_u(\bx_1), \ldots, \psi_u(\bx_n))$ and $\bxi = (\xi(\bx_1), \ldots, \xi(\bx_n))$.

The Poisson-GPD model's likelihood is slightly more challenging since it requires integration over all possible $\bx$, $\mathcal{X}$, say. Consequently \pkg{evgam} only currently considers models where integration is over time-dependent $\bx$, over which GEV parameters must be constant. Hence, consider $\{Y_t(\bx_i)\}$ for $i=1, \ldots, n$ and times $t = 1, \ldots, T$. The Poisson-GPD model's likelihood is
\begin{multline} \label{pplik} 
L(\bmu, \bpsi, \bxi) = \prod_{i = 1}^n \left[\exp \left\{ -n_{y}\left[1 + \xi(\bx_i)\left(\dfrac{y^{(r)}(\bx_i) - \mu(\bx_i)}{\psi(\bx_i)}\right)\right]^{-\frac{1}{\xi(\bx_i)}}\right\} \times \right. \\ \left. \prod_{t=1}^{r} \psi^{-1}\left[1 + \xi(\bx_i)\left(\dfrac{y^{(t)}(\bx_i) - \mu(\bx_i)}{\psi(\bx_i)}\right)\right]^{-\frac{1}{\xi(\bx_i)} - 1}\right]
\end{multline}
for time period $n_y$, $n$-vectors $\bmu$, $\bpsi$ and $\bxi$ and where $y^{(t)}(\bx)$, for $t=1, \ldots, T$, denote the order statistics of sample $y_1(\bx), \ldots, y_T(\bx)$ with $r < T$ chosen by the user. An example where $\mu(\bx)$, $\psi(\bx)$ and $\xi(\bx)$ vary with spatial locations is given in \S\ref{CO:pp}. 

The ALD is fitted to data relating to original random variables $\{Y(\bx_i)\}$ for $i = 1, \ldots, n$. Its likelihood is therefore
\[
L(\bu, \bsig) = \prod_{i=1}^{n} f_\text{ALD}\big(y(\bx_i); u(\bx_i), \sigma(\bx_i)\big),
\]
with with $\bu = (u(\bx_1), \ldots, u(\bx_n))$ and $\bsig = (\sigma(\bx_1), \ldots, \sigma(\bx_n))$.

Interval-censored data can also be fitted with \pkg{evgam}. Suppose $[y_-(\bx_i), y_+(\bx_i)]$ denotes the censoring interval of $y(\bx_i)$, a realization from $F( \, ; \cdot)$. Then the likelihood takes the form
\[L(\cdot) = \prod_{i=1}^n \big[F\big(y_+(\bx_i); \cdot\big) - F\big(y_-(\bx_i); \cdot\big)\big].\]

\subsection{Generalized additive modeling} \label{sec:models:gam}

The package \pkg{evgam} is primarily designed to allow nonstationarity in EVD parameters by assuming GAM forms in covariate $\bx$.

\subsubsection{Basis representations} \label{sec:models:basis}

GAM forms for EVD parameters rely on basis representations. Consider covariate $\bx$ and GEV parameters $\mu(\bx)$, $\psi(\bx)$ and $\xi(\bx)$. \pkg{evgam} relates parameters via fixed link functions to $\eta_*$, which has a basis representation. For the GEV, $\mu(\bx) = \eta_\mu(\bx)$, $\log \psi(\bx) = \eta_\psi(\bx)$ and $\xi(\bx) = \eta_\xi(\bx)$, where
\[
\eta_*(\bx) = \beta_{0} + \sum_{k=1}^{K} \sum_{d=1}^{D_k} \beta_{kd} b_{kd}(\bx)
\]
with $\beta_{kd}$ and $b_{kd}$ basis coefficients and functions, respectively. The upshot of the basis representation is that we can write $\eta_*(\bx) = {\bf x}^T {\bm \beta}$ where ${\bf x}^T$ is a row of a design matrix ${\bf X}$, which has elements determined by the choice of the $b_{kd}$ basis functions and $1 + \sum_{k=1}^K D_k$ columns, each of which corresponds to an element of ${\bm \beta}^T = (\beta_0, \beta_{11}, \ldots, \beta_{KD_K})$. The log link is used through \pkg{evgam} for any parameters with support $\mathds{R}^+$.

\subsubsection{Penalized likelihood} \label{sec:models:pen}

Various likelihoods were introduced in \S\ref{lik}. In general, consider data $\by = \{y_1, \ldots, y_n\}$ with corresponding covariates $\{\bx_1, \ldots, \bx_n\}$ and that estimating an EVD corresponds to estimating basis coefficients $\bbeta$. Hence each likelihood from \S\ref{lik} can be written $L(\bbeta)$ with log-likelihood $\ell(\bbeta)$.

To estimate EVD parameters a penalized log-likelihood of the form 
\[
\ell({\bm \beta}_{\bm \lambda}, {\bm \lambda}) = \ell({\bm \beta}_{\bm \lambda}) - \frac{1}{2}{\bm \beta}^T {\bf S}_{\bm \lambda} {\bm \beta},
\]
is considered for smoothing parameters ${\bm \lambda} = (\lambda_1, \ldots, \lambda_K)$, where ${\bf S}_{\bm \lambda}$ is a penalty matrix with elements determined by the chosen $b_{kd}$ basis functions. ${\bf S}_{\bm \lambda}$ may be written ${\bf S}_{\bm \lambda} = \sum_{k=1}^K \lambda_k {\bf S}_k$, where rows and columns of matrix ${\bf S}_k$ corresponding to $b_{k'd}$, $k' \neq k$, comprise zeros. Often the non-zero terms in the ${\bf S}_k$ matrices are non-overlapping. One contrary example is penalties constructed by tensor products \citep{deboor1978}; see \cite{wood2011} for fuller details.

\subsubsection{Restricted maximum likelihood} \label{reml}

Following \cite{wood2011}, ${\bm \beta}$ can be integrated out using Laplace's method, which results in a restricted log-likelihood of the form
\begin{equation} \label{remleq}
\ell({\bm \lambda}) = \ell(\hat {\bm \beta}_{\bm \lambda}, {\bm \lambda}) + \frac{1}{2} \log |{\bf S}_{\bm \lambda}|_+ - \frac{1}{2} \log |{\bf H}(\hat {\bm \beta}_{\bm \lambda})| + \text{constant},
\end{equation}
where $\hat {\bm \beta}_{\bm \lambda}$ maximizes $\ell({\bm \beta}_{\bm \lambda}, {\bm \lambda})$ for given ${\bm \lambda}$, ${\bf H}(\hat {\bm \beta}_{\bm \lambda}) = - \nabla\nabla^T \ell({\bm \beta}, {\bm \lambda})|_{{\bm \beta} = \hat {\bm \beta}_{\bm \lambda}}$ and $|{\bf S}_{\bm \lambda}|_+$ denotes the product of positive eigenvalues of matrix ${\bf S}_{\bm \lambda}$. Optimal smoothing parameters, $\hat {\bm \lambda}$, can be found by numerically maximising $\ell({\bm \lambda})$, which is typically best performed through Newton or quasi-Newton methods, as implemented by \pkg{evgam}. Fitting a model therefore involves inner iterations, for given ${\bm \lambda}$, which give $\hat {\bm \beta}_{\bm \lambda}$, and outer iterations, which give $\hat {\bm \lambda}$. 

\section[Features of evgam]{Features of \pkg{evgam}} \label{sec:evgam}

\subsection[Function evgam()]{Function \fct{evgam}} \label{evgam}

\subsubsection{Basic use}

The package \pkg{evgam} mainly relies on its eponymous function \fct{evgam}. Its main arguments are 
\begin{Code}
evgam(formula, data, family)
\end{Code}

Typically \code{formula} is a list comprising formulae: one formula compatible with \fct{mgcv::s} for each EVD parameter. Hence, see the help for \fct{mgcv::s} for details of its use. If a single formula is supplied, it is repeated for each EVD parameter so that the same form is assumed for each parameter. Use of \code{data} is the same as for, e.g., \fct{lm}. Interval-censored data can also be handled with \code{formula}. Supplying \code{cens(left, right)} as the response states that \code{data$left} are \code{data$right} variables giving lower and upper ends of the censoring interval, respectively. Any response data for which \code{data$left} and \code{data$right} are equal are treated as uncensored. (Note that left- and right-censored data can be handled with sufficiently high and low lower and upper interval ranges, respectively.) An example fitting the GPD to censored data is given in \S\ref{CO:cgpd}.

The default \code{family} is \code{"gev"}, which corresponds to the GEV distribution. GPD, Poisson-GPD and ALD models are specified with \code{"gpd"}, \code{"pp"} and \code{"ald"}, respectively. \pkg{evgam} also supports fitting of exponential, \code{"exponential"}, Weibull, \code{"weibull"} and Gaussian, \code{"gauss"}, distributions. 

For the ALD, the quantile to be estimated must be given: supplying \code{ald.args = list(tau = 0.9)}, for example,  gives an estimate of the 0.9 quantile. For the point process model, the time period under study and the number of order statistics to use are required: supplying \code{pp.args = list(ny = 30, r = 50)} specifies a 30-period time period, e.g. 30 years, if parameters representative of annual maxima are sought, and 50 order statistics. Note that \code{r = -1} uses all order statistics. Fitting ALD and Poisson-GPD models is demonstrated in \S\ref{FC:dy} and \S\ref{CO:pp}, respectively. \S\ref{CO:pp} also demonstrates how \code{pp.args$id} may be used to specify partitions of \code{data} over which integration is not required.

\subsubsection{Additional options}

The default values used by \pkg{evgam} are designed to be robust. In some circumstances, however, changes to some arguments' default values may improve performance. First consider \code{trace}, which accepts 0 (default), 1, 2 or -1; increasing numbers report more on optimization iterations, and -1 reports nothing. \code{trace} can be useful for ensuring that inner and/or outer iterations have converged. There are two arguments that may improve speed for large datasets. First, \code{maxdata} specifies the maximum number of rows in \code{data} that will be used in model fitting: if \code{nrow(data) > maxdata} then \code{maxdata} rows of \code{data} are sampled without replacement. Second, \code{maxspline} specifies the maximum number of rows in \code{data} that are supplied to \fct{mgcv::s} to create bases; all rows of \code{data} are then used for fitting unless \code{maxdata > maxspline} is also invoked. Initial values for $\rho_k = \log \lambda_k$, $k=1, \ldots, K$, are supplied with \code{rho0}; \pkg{evgam}'s default is $\lambda_k = 1$ for all $k$. Providing a scalar specifies the same initial value for each $\lambda_k$, whereas a vector of length $K$ allows different initial values. Argument \code{inits} allows initial values for ${\bm \beta}_\lambda$ to be specified in various ways, such as subsets of ${\bm \beta}_\lambda$. Argument \code{outer} specifies how the restricted log-likelihood of Eq. \eqref{remleq} is optimized: the default, \code{"BFGS"}, uses the BFGS quasi-Newton method; \code{"Newton"} uses Newton's method; and \code{"FD"} uses BFGS with finite-difference approximations to the gradient of Eq. \eqref{remleq} w.r.t. each $\rho_k$. See \fct{evgam}'s help file for details of its other options.

\subsection[Function qev()]{Function \fct{qev}} \label{qev}

Also included in \pkg{evgam} is \fct{qev} for quantiles of EVDs. It solves $F_\text{ann}(z_p) = p$, numerically where necessary, for $z_p$. Its arguments are
\begin{Code}
qev(p, loc, scale, shape, m = 1, alpha = 1, theta = 1, family, tau = 0)
\end{Code}
In the above \code{p} is $p$ in $F_\text{ann}(z_p) = p$, and \code{loc}, \code{scale} and \code{shape} are an EVD's location, scale and shape parameters, respectively. In terms of \S\ref{sec:models:rlns}, \code{m} corresponds to $n_y$, \code{alpha} to $w(\,)$, \code{theta} to $\theta$, \code{family} is that supplied to \fct{evgam}, and \code{tau} corresponds to $1 - \zeta$. 


\section{Illustrations} \label{sec:illustrations}

Illustrations for \pkg{evgam} are given below. All require \pkg{evgam} to be loaded, which is done with
\begin{Schunk}
\begin{Sinput}
R> library(evgam)
\end{Sinput}
\end{Schunk}

\subsection{Spatial modeling: Colorado precipitation} \label{CO}

To illustrate the key functionality of \pkg{evgam} the dataset \code{COprcp} 
will be used, which contains daily precipitation amounts, \code{prcp}, in mm on day 
\code{date} at locations identified by \code{meta_row} for part of Colorado, US. (This was the
domain studied in \cite{cooley2007}.) Each location's metadata corresponds to a row in \code{COprcp_meta}.

\subsubsection{The COprcp data} \label{CO:data}

The data can be loaded and conjoined with the metadata using
\begin{Schunk}
\begin{Sinput}
R> data(COprcp)
R> COprcp <- cbind(COprcp, COprcp_meta[COprcp$meta_row,])
\end{Sinput}
\end{Schunk}
\begin{figure}[t!]
\centering
\includegraphics[height=.4\textheight, width=.2\textheight]{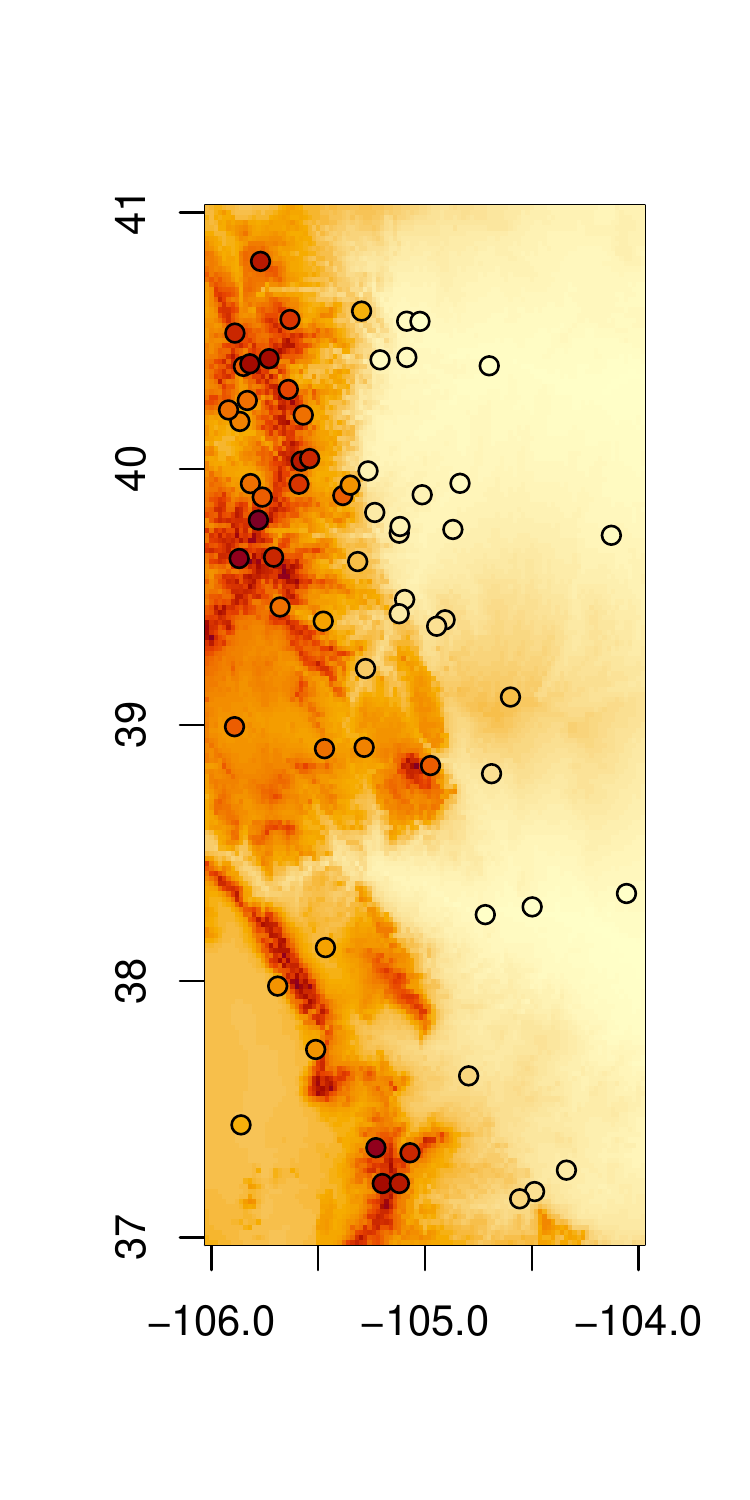}
\caption{\label{fig:COelev} Gridded and station-based elevation data for study region.}
\end{figure}
The dataset \code{COprcp} also includes \code{COelev}, gridded elevation data
for the study region. A plot of gridded elevations (Fig. \ref{fig:COelev}) can be obtained with
\begin{Schunk}
\begin{Sinput}
R> brks <- pretty(COelev$z, 50)
R> cols <- hcl.colors(length(brks) - 1, "YlOrRd", rev = TRUE)
R> image(COelev, breaks = brks, col = cols, asp = 1)
\end{Sinput}
\end{Schunk}
The function \fct{colplot} is included in \pkg{evgam} to plot points that are colored according to a variable. Station elevations can be superimposed on the gridded elevations of Fig. \ref{fig:COelev} with
\begin{Schunk}
\begin{Sinput}
R> colplot(COprcp_meta$lon, COprcp_meta$lat, COprcp_meta$elev, 
+          breaks = brks, palette = cols, add = TRUE)
\end{Sinput}
\end{Schunk}
Before fitting any models, a \code{data.frame} for plotting, \code{COprcp_plot}, is created using
\begin{Schunk}
\begin{Sinput}
R> COprcp_plot <- expand.grid(lon = COelev$x, lat = COelev$y)
R> COprcp_plot$elev <- as.vector(COelev$z)
\end{Sinput}
\end{Schunk}
Subsequent models will use elevation as a covariate, so it has been included in \code{COprcp_plot}. 
Coordinate and covariate names match those in \code{COprcp_meta}.

\subsubsection{GEV model} \label{CO:gev}

First we model annual maxima using the GEV distribution, introduced in \S\ref{sec:models:gev}. 
This model will be implemented by creating a \code{data.frame} comprising annual maxima at each station.
Since \code{date} is of class \code{"Date"}, this can be done with
\begin{Schunk}
\begin{Sinput}
R> COprcp$year <- format(COprcp$date, "%Y")
R> COprcp_gev <- aggregate(prcp ~ year + meta_row, COprcp, max)
\end{Sinput}
\end{Schunk}
which aggregates over \code{meta_row}, i.e., over the station IDs, and then the metadata can be added to \code{COprcp_gev} with
\begin{Schunk}
\begin{Sinput}
R> COprcp_gev <- cbind(COprcp_gev, COprcp_meta[COprcp_gev$meta_row,])
\end{Sinput}
\end{Schunk}
The next step is to provide formulae for smooths to pass to \fct{mgcv::s}. A spatial model will be fitted
that allows spatial variation in the GEV's location and scale parameters. Spatial variation is achieved with thin plate
regression splines, which are \fct{mgcv::s}'s default. The basis dimension, \code{k}, has been specified
to differ with GEV parameter. The GEV's shape parameter is assumed constant. The value of \code{k}
caps a smooth's degrees of freedom, and hence, in some sense, its ultimate wiggliness. In practice, \code{k} should 
be chosen larger than a smooth's expected degrees of freedom so that the smoothing parameters control the effective degrees
of freedom. The GEV's location parameter also includes a smooth in \code{elev}, station elevation. This is
specified as a cubic regression spline, \code{bs = "cr"}, with \code{k} left at its default. The smooths for all GEV parameters are then specified with
\begin{Schunk}
\begin{Sinput}
R> fmla_gev <- list(prcp ~ s(lon, lat, k = 30) + s(elev, bs = "cr"), 
+    ~ s(lon, lat, k = 20), ~ 1)
\end{Sinput}
\end{Schunk}
To fit the model we issue
\begin{Schunk}
\begin{Sinput}
R> m_gev <- evgam(fmla_gev, COprcp_gev, family = "gev")
\end{Sinput}
\end{Schunk}
(but could have omitted \code{family = "gev"} above since it is \fct{evgam}'s default).

Having fitted the model, it is sensible to check whether smooths are necessary, and if so whether they are well specified.
This can be done through \fct{summary} with
\begin{Schunk}
\begin{Sinput}
R> summary(m_gev)
\end{Sinput}
\begin{Soutput}
** Parametric terms **

location
            Estimate Std. Error t value Pr(>|t|)
(Intercept)    28.56       0.26  111.89   <2e-16

logscale
            Estimate Std. Error t value Pr(>|t|)
(Intercept)     2.24       0.02  118.07   <2e-16

shape
            Estimate Std. Error t value Pr(>|t|)
(Intercept)     0.08       0.02    5.08 1.92e-07

** Smooth terms **

location
             edf max.df Chi.sq Pr(>|t|)
s(lon,lat) 19.27     29 178.23   <2e-16
s(elev)     5.19      9  19.39  0.00139

logscale
             edf max.df Chi.sq Pr(>|t|)
s(lon,lat) 13.94     19 211.15   <2e-16
\end{Soutput}
\end{Schunk}
The necessity of smooths can be checked through $p$-values. These are all $\ll 0.01$, indicating that they are beneficial. All one- or two-dimensional smooths can be then viewed with \fct{plot}, i.e., 
\begin{Schunk}
\begin{Sinput}
R> plot(m_gev)
\end{Sinput}
\end{Schunk}
\begin{figure}[t!]
\centering
\includegraphics[scale=0.5]{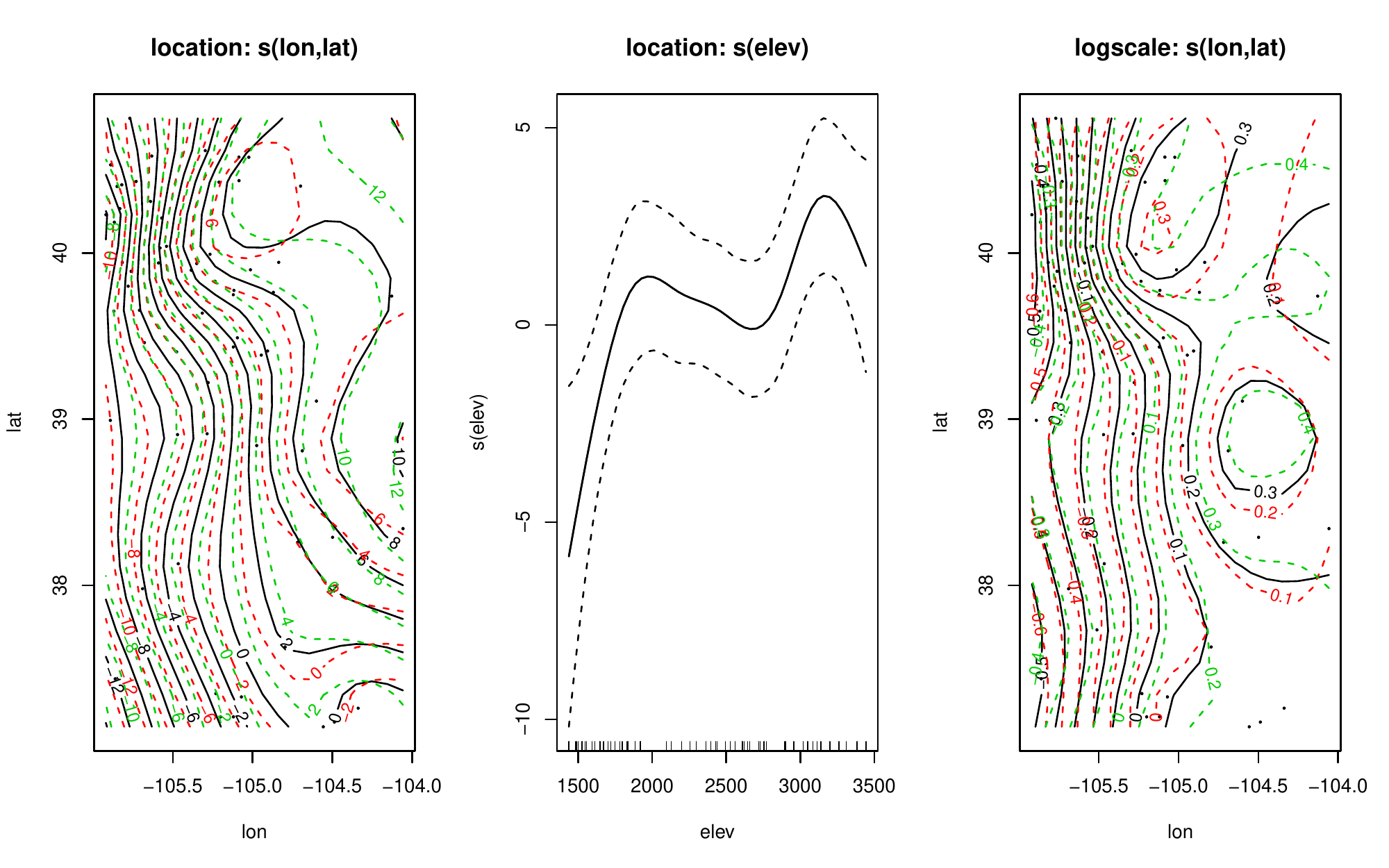}
\caption{\label{fig:m_gev} Output of \code{plot(m\_gev)} for Colorado precipitation annual maxima.}
\end{figure}
which is shown in Fig. \ref{fig:m_gev}. Often predictions are sought from a fitted model. These are achieved with \fct{predict}. Predictions for the GEV's three parameters for \code{COprcp_plot} can be obtained with
\begin{Schunk}
\begin{Sinput}
R> gev_pred <- predict(m_gev, COprcp_plot, type = "response")
R> head(gev_pred)
\end{Sinput}
\begin{Soutput}
  location    scale      shape
1 12.79505 5.344232 0.07941214
2 13.09313 5.422900 0.07941214
3 13.38081 5.503366 0.07941214
4 13.67835 5.585679 0.07941214
5 13.97230 5.669885 0.07941214
6 14.27654 5.756028 0.07941214
\end{Soutput}
\end{Schunk}
where \fct{head} is used here (and later) to suppress estimates for all but the first six rows of \fct{predict}'s output. Note that \code{type = "response"} is used to predict parameters on their original scale, similarly to \fct{predict.glm}. Hence \code{gev_pred} is a three-column \code{data.frame} with columns for the GEV location, scale and shape parameters, respectively. Predictions can be viewed with \fct{image} using a few lines of code (omitting the constant shape parameter), such as 
\begin{Schunk}
\begin{Sinput}
R> for (i in 1:2) {
+    plot.list <- COelev
+    plot.list$z[] <- gev_pred[,i]
+    image(plot.list, asp = 1)
+    title(paste("GEV", names(gev_pred)[i]))
+  }
\end{Sinput}
\end{Schunk}
\begin{figure}[t!]
\centering
\includegraphics[height=.58\textwidth, width=.325\textwidth, page=1]{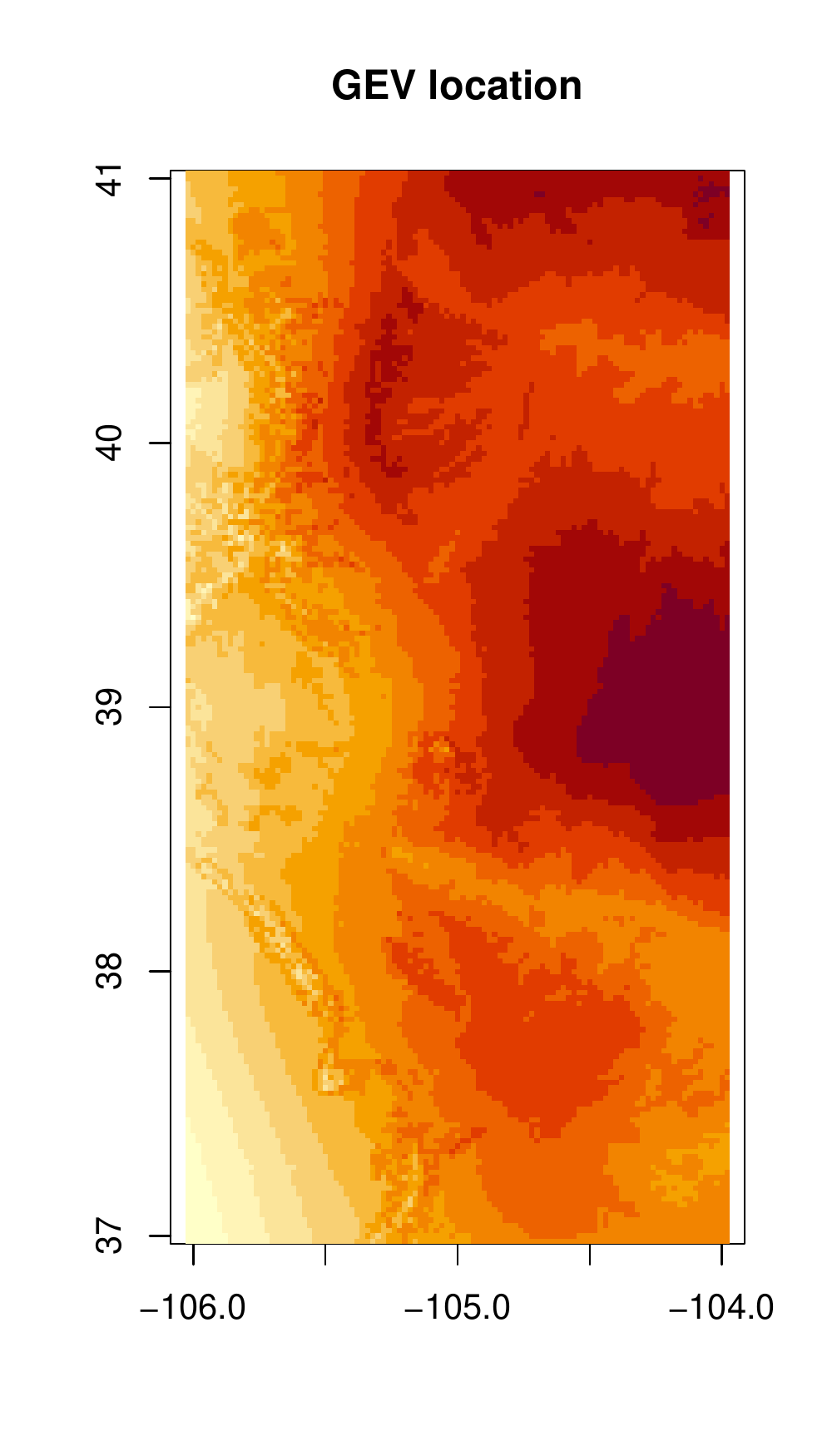}
\includegraphics[height=.58\textwidth, width=.325\textwidth, page=2]{evgam_v3-gev_pred_plot.pdf}
\includegraphics[height=.58\textwidth, width=.325\textwidth]{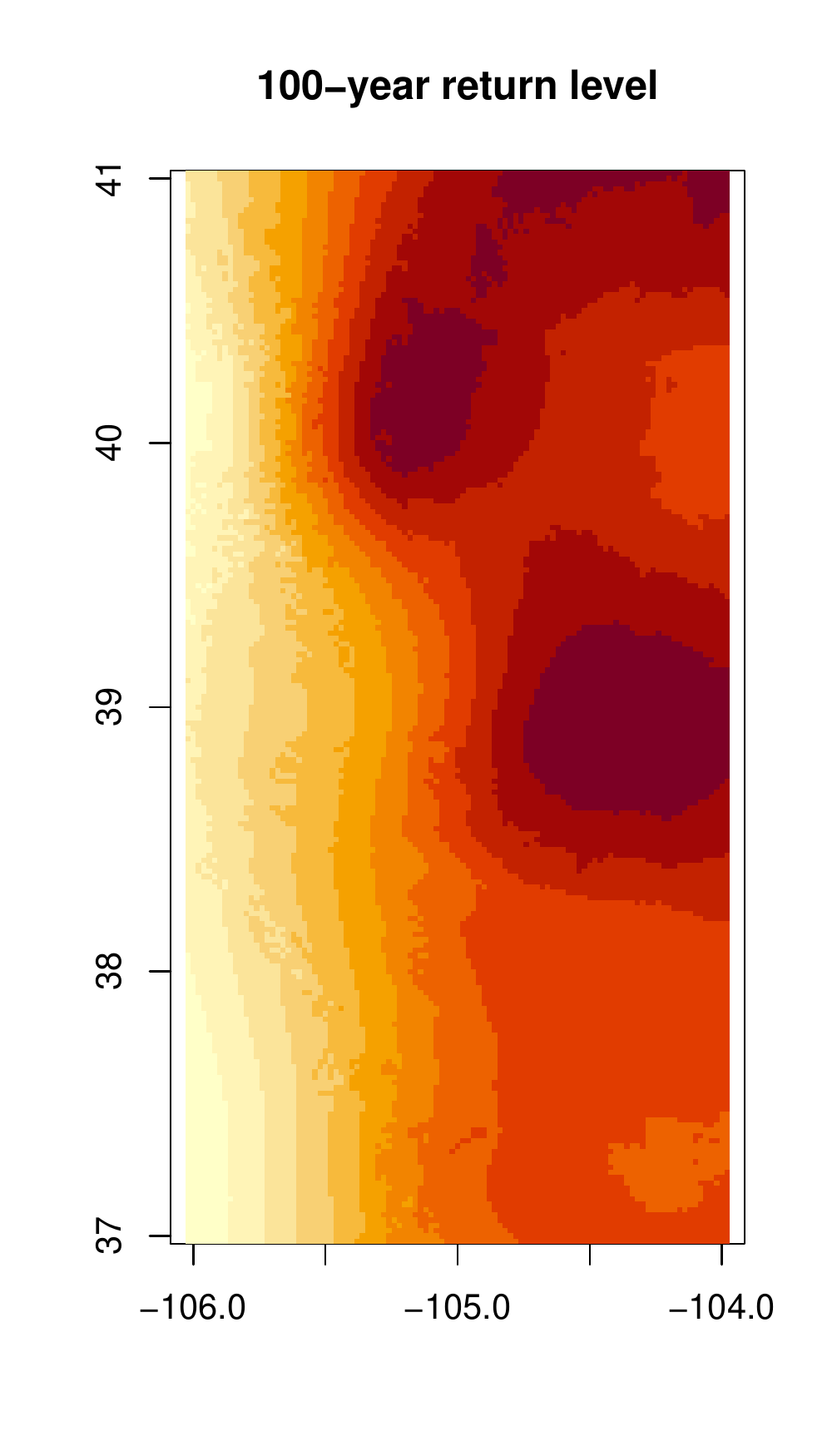}
\caption{\label{fig:plot_gev_pred} Plots of GEV parameter estimates for Colorado precipitation annual maxima and of the 100-year return level estimate.}
\end{figure}

which is shown in Fig. \ref{fig:plot_gev_pred}. Lastly, the 100-year return level for the locations in \code{COprcp_plot} can be estimated. This is an estimate of the 0.99 quantile of the distribution of the annual maximum for \emph{each} location and achieved with
\begin{Schunk}
\begin{Sinput}
R> gev_rl100 <- predict(m_gev, COprcp_plot, prob = 0.99)
R> head(gev_rl100)
\end{Sinput}
\begin{Soutput}
    q:0.99
1 42.47033
2 43.20524
3 43.93974
4 44.69434
5 45.45587
6 46.23844
\end{Soutput}
\end{Schunk}
and plotted using
\begin{Schunk}
\begin{Sinput}
R> rl100 <- COelev
R> rl100$z[] <- gev_rl100[,1]
R> image(rl100, asp = 1)
R> title("100-year return level")
\end{Sinput}
\end{Schunk}
which is also shown in Fig. \ref{fig:plot_gev_pred}. Uncertainty estimates, in particular for return levels, are covered in \S\ref{sec:uq}.

\subsubsection{GPD model} \label{CO:gpd}

The GPD is used to model excesses of a high threshold. Here, following \cite{cooley2007}, the threshold is set at 11.4mm using
\begin{Schunk}
\begin{Sinput}
R> threshold <- 11.4
\end{Sinput}
\end{Schunk}
To fit the GPD only threshold exceedances are considered. Setting excesses corresponding to non-exceedances as \code{NA} ensures that only exceedances are modeled, which is done using  
\begin{Schunk}
\begin{Sinput}
R> COprcp$excess <- COprcp$prcp - threshold
R> COprcp$excess[COprcp$excess <= 0] <- NA
\end{Sinput}
\end{Schunk}
A similar formula, in terms of smooths, is used for the GPD model as was used for the GEV model, although this model comprises only two parameters and a non-constant shape parameter is allowed. A smooth with \code{elev} is included for the GPD's scale parameter, which is partly motivated by use of a constant threshold. A varying threshold model is given in \S\ref{FC:dy}. The GPD model is fitted with
\begin{Schunk}
\begin{Sinput}
R> fmla_gpd <- list(excess ~ s(lon, lat, k = 20) + s(elev, bs = "cr"), 
+    ~ s(lon, lat, k = 15))
R> m_gpd <- evgam(fmla_gpd, COprcp, family = "gpd")
\end{Sinput}
\end{Schunk}
Summaries, plots and predictions can be produced for \code{m_gpd} as demonstrated above for \code{m_gev}, and so are not demonstrated again. Using \code{predict(..., prob = ...)} if \code{family = "gpd"} uses Eq. \eqref{gpdrl}. The example of \S\ref{FC:dy} demonstrates return level estimation in the presence of dependence.

\subsubsection{Poisson-GPD model} \label{CO:pp}

For the point process model, following \S\ref{lik}, our data will be consdiered realizations of $\{Y_t(\bx)\}$ for location $\bx$ in region $\mathcal{X}$ and time $t = 1, \ldots, T$. Hence covariate $\bx$ is not time-dependent, and log-likelihood \eqref{pplik} is used. If different locations have different $T$, $n_y(\bx)$ should be used in log-likelihood \eqref{pplik}. \pkg{evgam} facilitates that by allowing vector \code{pp.args$ny}. Note that \code{names(pp.args$ny)} must match unique \code{pp.args$id} to ensure that correct $n_y(\bx)$ and $\{Y_t(\bx)\}$ coincide.

Different stations in \code{COprcp} are identified by variable \code{id}. We want to assume a constant point process rate for a given \code{id}. We do this by setting \code{pp.args$id} to \code{"id"}. (Double use of `id' is a coincidence.) For this model \code{fmla_gev} is re-used and then \fct{evgam} called with
\begin{Schunk}
\begin{Sinput}
R> pp_args <- list(id = "id", ny = 30, r = 45)
R> m_pp <- evgam(fmla_gev, COprcp, family = "pp", pp.args = pp_args)
\end{Sinput}
\end{Schunk}
In the above the 45 largest observations at each station are used, and 30 periods of observation at each station is specified. \code{COprcp} comprises 30 years' data (aside from a few missing values) at each station; hence the Poisson-GPD model's GEV parameter estimates will represent the distribution of the annual maximum.

Summaries, plots and predictions can be produced for \code{m_pp} similarly to \code{m_gev}, and so are again omitted for brevity. Note that $r$-largest order statistics at a given station may exhibit dependence similarly to threshold excesses and so the same considerations for \code{predict(..., prob = ...)} as for the GPD apply.

\subsubsection{Censored response data and tensor products: GPD model revisited} \label{CO:cgpd}

\cite{cooley2007} allude to precipitation being recorded with relatively little precision. Sometimes such data may want to be treated as censored. For example, continuous data recorded to the nearest integer, $x$, say, could be treated as interval-censored on $[x - 0.5, x + 0.5)$. Alternatively, measurement $x$ might be given with stated tolerance $\delta$, i.e., $x \pm \delta$, so that the response should be treated as interval-censored on $[x - \delta, x + \delta]$. \cite{cooley2007} state that some precipitation values were recorded to the nearest tenth of an inch, or $\sim 2.5$mm. One option for setting up the censoring interval is
\begin{Schunk}
\begin{Sinput}
R> delta <- 2.5
R> COprcp$lo <- pmax(COprcp$excess - delta, 1e-6)
R> COprcp$hi <- COprcp$excess + delta
\end{Sinput}
\end{Schunk}
Tensor products, e.g., \cite{deboor1978} and \cite{wood2006b}, can be used to specify interactions between smooths. For example, instead of a thin plate regression spline, a two-dimensional smooth can be formed through the tensor product of two one-dimensional smooths. The earlier GPD formula is modified for interval-censored response data and spatial smooths formed from two cubic regression splines with
\begin{Schunk}
\begin{Sinput}
R> fmla_gpd_cens <- list(cens(lo, hi) ~ te(lon, lat, k = c(6, 8)) +
+    s(elev, bs = "cr"), ~ te(lon, lat, k = c(6, 8)))
\end{Sinput}
\end{Schunk}
which specifies rank 6 and rank 8 cubic regression splines for longitude and latitude (a choice based on the tall rectangular shape of the domain). The GPD is then fit as above, but with a new \code{formula}, and plotted with
\begin{Schunk}
\begin{Sinput}
R> m_gpd_cens <- evgam(fmla_gpd_cens, COprcp, family = "gpd")
R> plot(m_gpd_cens)
\end{Sinput}
\end{Schunk}
\begin{figure}[t!]
\centering
\includegraphics[scale=0.5]{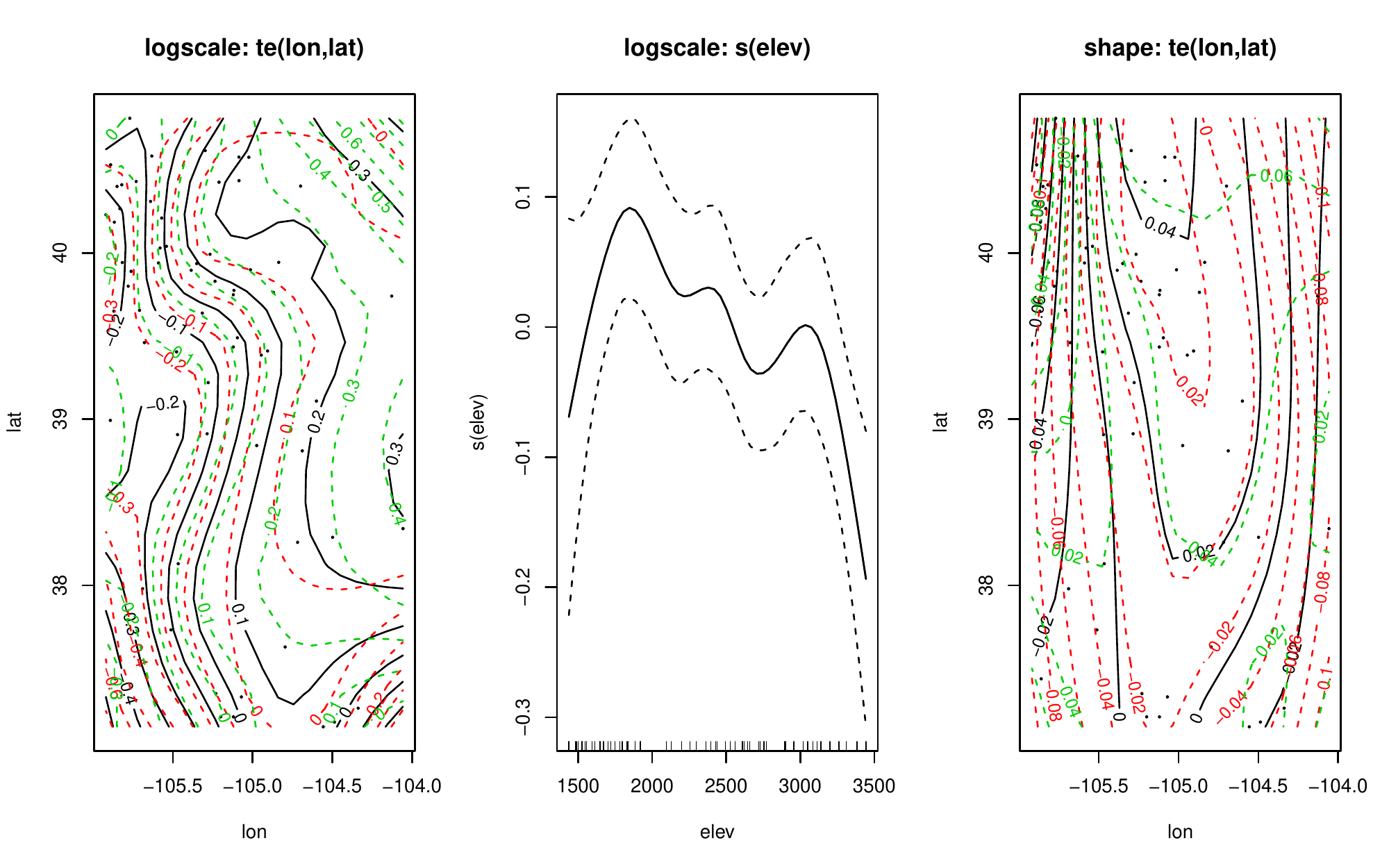}
\caption{\label{fig:cens_gpd} Output of \code{plot(m\_gpg\_cens)} for Colorado precipitation threshold exceedances treated as censored with spatial smooths formed from tensor products.}
\end{figure}
which is shown in Figure \ref{fig:cens_gpd}.

\subsection{Temporal modeling: Fort Collins temperatures}

This example considers \code{FCtmax}, a data frame comprising daily maximum temperatures, \code{tmax}, in degrees Celsius at Fort Collins, Colorado, US. The data cover 1st January 1970 to 31st December 2019. There are 95 missing values during this period. Two different approaches to assuming that the distribution of extreme temperatures changes throughout the year are considered. The aim is to estimate the 100-year return level.

The data are loaded using
\begin{Schunk}
\begin{Sinput}
R> data(FCtmax)
\end{Sinput}
\end{Schunk}
\subsubsection{GEV model for monthly maxima} \label{FC:mn}

The first model uses monthly maxima and its first step is to identify the monthly maxima. Dates are identified by \code{date}, of class \code{"Date"}, so years and months are obtained with
\begin{Schunk}
\begin{Sinput}
R> FCtmax$year <- format(FCtmax$date, "%Y")
R> FCtmax$month <- format(FCtmax$date, "%m")
\end{Sinput}
\end{Schunk}
and then \fct{aggregate} can be used to find the monthly maxima with
\begin{Schunk}
\begin{Sinput}
R> FCtmax_mnmax <- aggregate(tmax ~ year + month, FCtmax, max)
\end{Sinput}
\end{Schunk}
There are various ways to proceed. Here \code{FCtmax_mnmax} is separated by month with \fct{split}, i.e., 
\begin{Schunk}
\begin{Sinput}
R> FCtmax_mn <- split(FCtmax_mnmax, FCtmax_mnmax$month)
\end{Sinput}
\end{Schunk}
which gives a \code{list} of \code{data.frame}s, each of which comprises monthly maxima over years for a given month. 
%

GEV parameter estimates for each month's maxima are obtained with
\begin{Schunk}
\begin{Sinput}
R> fmla_simple <- list(tmax ~ 1, ~ 1, ~ 1)
R> gev_fits <- lapply(FCtmax_mn, evgam, formula = fmla_simple, family = "gev")
R> gev_pars <- sapply(gev_fits, coef)
\end{Sinput}
\end{Schunk}
where \code{fmla_simple} specifies that for a given month all GEV parameters are constant.

The function \fct{qev} is then used to estimate the 100-year return level using Eq. \eqref{Fann_gev} from \S\ref{sec:models:rlns}. This requires the weights $w(\bx_i)$ for $i=1, \ldots, 12$. These are simply
\begin{Schunk}
\begin{Sinput}
R> weights <- (1/365.25) * c(31, 28.25, 30)[c(1, 2, 1, 3, 1, 3, 1, 1, 3, 1, 3, 1)]
\end{Sinput}
\end{Schunk}
and are supplied to \fct{qev}, documented in \S\ref{qev}, using
\begin{Schunk}
\begin{Sinput}
R> rl_100_gev <- qev(0.99, gev_pars[1,], exp(gev_pars[2,]), gev_pars[3,], 
+    m = 12, alpha = weights, family = "gev")
\end{Sinput}
\end{Schunk}
This gives a 100-year return level estimate of 39.37$^\circ$C.

\subsubsection{GPD model for daily threshold excedances} \label{FC:dy}

\begin{figure}[t!]
\centering
\includegraphics[height=.35\textheight, width=.7\textwidth]{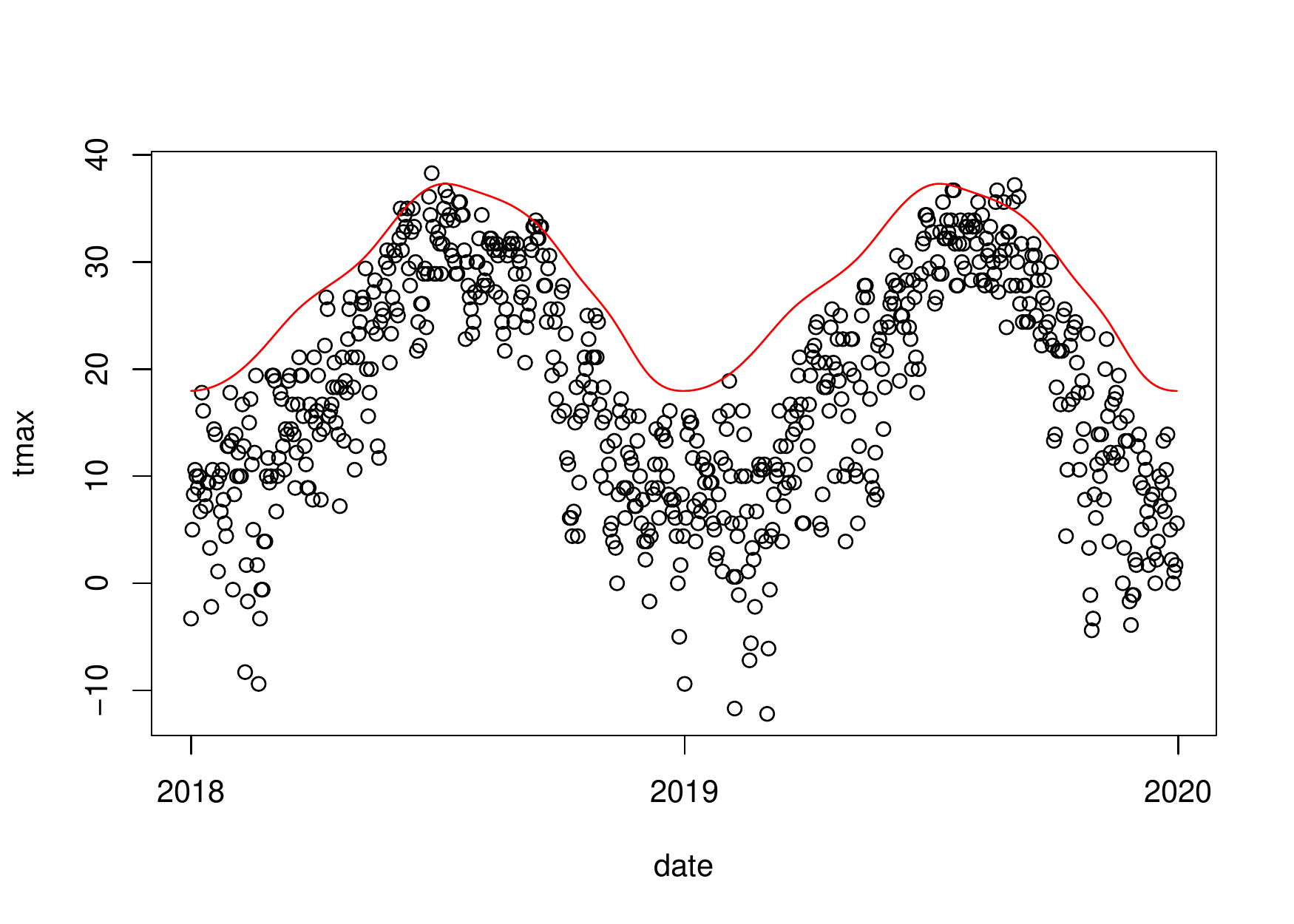}
\caption{\label{fig:FC:thresh} Daily maximum temperatures at Fort Collins for 2018 and 2019 with a cyclic estimate of the 99th percentile superimposed.}
\end{figure}

What is an extreme temperature at one time of the year is different from that occurring at another time of the year. As a result, extreme values are now defined as exceedances of a time-varying threshold. The threshold itself is estimated as the 99th percentile by quantile regression. Hence $\zeta = 0.01$, given Eq. \eqref{Fann_gpd} from \S\ref{sec:models:rlns}, so we set
\begin{Schunk}
\begin{Sinput}
R> zeta <- 0.01
\end{Sinput}
\end{Schunk}
A threshold estimate that varies over a course of a year and that is the same and continuous from year to year is sought. This is achieved through a \emph{cyclic} cubic regression spline, specified with \code{bs = "cc"} in \fct{mgcv::s}. The variable \code{cyc} is therefore created using
\begin{Schunk}
\begin{Sinput}
R> FCtmax$cyc <- as.integer(FCtmax$date) %% 365.25
\end{Sinput}
\end{Schunk}
The formula for the model is specified, and then the model fitted, using
\begin{Schunk}
\begin{Sinput}
R> FC_fmla_ald <- list(tmax ~ s(cyc, bs = "cc", k = 15),  ~ s(cyc, bs = "cc"))
R> FC_ald <- evgam(FC_fmla_ald, FCtmax, family = "ald", 
+    ald.args = list(tau = 1 - zeta))
\end{Sinput}
\end{Schunk}
Variables for the estimated threshold, \code{threshold}, and resulting excesses, \code{excess}, are added to \code{FCtmax} using
\begin{Schunk}
\begin{Sinput}
R> FCtmax$threshold <- predict(FC_ald)$location
R> FCtmax$excess <- FCtmax$tmax - FCtmax$threshold
R> FCtmax$excess[FCtmax$excess <= 0] <- NA
\end{Sinput}
\end{Schunk}
It is quite useful to superimpose the threshold estimate on a scatter plot of the data, which is shown in Fig. \ref{fig:FC:thresh} for 2018 and 2019's data, and obtained using
\begin{Schunk}
\begin{Sinput}
R> use <- FCtmax$year %in% c("2018", "2019")
R> plot(FCtmax[use, c("date", "tmax")])
R> lines(FCtmax[use, c("date", "threshold")], col = "red")
\end{Sinput}
\end{Schunk}
Having established that the estimated threshold is satisfactory, its excesses are modeled as GPD realizations with
\begin{Schunk}
\begin{Sinput}
R> FC_fmla_gpd <- list(excess ~ s(cyc, bs = "cc", k = 15), ~ 1)
R> FC_gpd <- evgam(FC_fmla_gpd, FCtmax, family = "gpd")
\end{Sinput}
\end{Schunk}
which assumes a cyclic form for the scale parameter and a constant shape parameter. Note that setting non-exceedances to \code{NA} earlier ensured they were ignored by \fct{evgam}.

It is not reasonable to assume that these excesses of the threshold are independent. Hence to estimate the 100-year return level using $F_\text{ann}$ for the GPD's nonstationary case, introduced in \S\ref{sec:models:rlns}, allowance needs to be made for clustering: i.e., an estimate of the extremal index, $\theta$, is needed. The function \fct{extremal} is used to give an estimate based on the moment-based estimator of \cite{ferro2003}. This is implemented with
\begin{Schunk}
\begin{Sinput}
R> theta <- extremal(!is.na(FCtmax$excess), FCtmax$date)
\end{Sinput}
\end{Schunk}
where \code{FCtmax$date} is supplied to allow the missing values in \code{FCtmax$tmax} to be identified. This gives an extremal index estimate of $0.498$, corresponding to an average cluster size, defined in terms of grouped threshold exceedances, of $2.01$ days. 

To estimate the 100-year return level, finite values of the continuous variable \code{cyc} need to be chosen. We could simply choose \code{1:365}. There may, however, be occasions when the numerical estimate is computationally expensive. If the cyclic form is fairly smooth, fewer points can then be used. This is demonstrated here with the use of 50 points instead. A \code{data.frame} of 50 \code{cyc} values is created using
\begin{Schunk}
\begin{Sinput}
R> rl_df <- data.frame(cyc=seq(0, 365.25, l = 51)[-1])
R> rl_df$threshold <- predict(FC_ald, rl_df, type = "response")$location
R> rl_df[,c("psi", "xi")] <- predict(FC_gpd, rl_df, type = "response")
\end{Sinput}
\end{Schunk}
and then \fct{qev} used to estimate the 100-year return level with
\begin{Schunk}
\begin{Sinput}
R> rl_100_gpd <- qev(0.99, rl_df$threshold, rl_df$psi, rl_df$xi, m = 365.25, 
+    theta = theta, family = "gpd", tau = 1 - zeta)
\end{Sinput}
\end{Schunk}
which gives a 100-year return level estimate of 39.1$^\circ$C. 

Return level estimates corresponding to monthly maxima can also be obtained with this model. For example, using \code{rl_df <- data.frame(cyc = 1:31)} above would use 31 \code{cyc} points, i.e., each day in January, to estimate the 100-January return level.

\subsection{Uncertainty estimation} \label{sec:uq}

The above Colorado precipitation and Fort Collins temperature examples are used in this section to demonstrate the various options for uncertainty estimation available with \pkg{evgam}.

\subsubsection{Standard errors for EVD parameters}

First consider uncertainty estimates for parameters of an EVD. The GEV model of \S\ref{CO:gev} will be used for demonstration. The key function here is \fct{predict} using argument \code{se.fit = TRUE}. Standard error estimates for GEV parameters estimated for each row of \code{COprcp_plot} using \code{m_gev} can be obtained with
\begin{Schunk}
\begin{Sinput}
R> gev_pred <- predict(m_gev, COprcp_plot, type = "response", se.fit = TRUE)
R> head(gev_pred$se.fit)
\end{Sinput}
\begin{Soutput}
  location    scale      shape
1 1.994659 1.040021 0.01565573
2 1.959301 1.039746 0.01565573
3 1.926856 1.039144 0.01565573
4 1.890763 1.038210 0.01565573
5 1.855742 1.036937 0.01565573
6 1.815885 1.035324 0.01565573
\end{Soutput}
\end{Schunk}
which has shown just the standard error estimates, stored as \code{se.fit}.

\subsubsection{Standard errors for return levels}

Uncertainty estimates for return levels can also be produced. These rely on the Delta method and are achieved with 
\begin{Schunk}
\begin{Sinput}
R> gev_rl100_pred <- predict(m_gev, COprcp_plot, prob = c(0.95, 0.99), 
+    se.fit = TRUE)
R> head(gev_rl100_pred$se.fit)
\end{Sinput}
\begin{Soutput}
    q:0.95   q:0.99
1 4.517385 6.742178
2 4.495358 6.726219
3 4.473734 6.709576
4 4.449786 6.690467
5 4.425445 6.670132
6 4.398455 6.647173
\end{Soutput}
\end{Schunk}
which has shown the standard error estimates for the 0.95 and 0.99 quantiles of the GEV distribution.

\subsubsection{Simulation of EVD parameters and return levels}

Sampling distributions of EVD parameters or return levels can be skewed. Standard errors will not capture this. The \fct{simulate} function generates samples of parameters or return levels. \code{nsim = 5} samples for each GEV parameter from the model of \S\ref{CO:gev} for each row of \code{COprcp_plot} are generated using
\begin{Schunk}
\begin{Sinput}
R> gev_sim <- simulate(m_gev, nsim = 5, newdata = COprcp_plot, type = "response")
R> lapply(gev_sim, head)
\end{Sinput}
\begin{Soutput}
$location
      [,1]     [,2]      [,3]     [,4]     [,5]
1 13.12826 15.85528  8.156803 15.76731 8.612907
2 13.46851 16.13148  8.592327 16.08122 8.823714
3 13.79065 16.37516  9.008439 16.39294 9.065781
4 14.13144 16.61405  9.442720 16.73099 9.281043
5 14.46592 16.83062  9.870683 17.07582 9.508744
6 14.82116 17.02257 10.315125 17.46658 9.706280

$scale
      [,1]     [,2]     [,3]     [,4]     [,5]
1 7.518972 6.091424 4.865701 5.804763 5.816610
2 7.589078 6.154513 4.938312 5.884048 5.891826
3 7.660088 6.218500 5.012996 5.964711 5.967784
4 7.732016 6.283405 5.089832 6.046777 6.044484
5 7.804873 6.349246 5.168892 6.130272 6.121930
6 7.878670 6.416039 5.250249 6.215221 6.200126

$shape
        [,1]       [,2]       [,3]     [,4]       [,5]
1 0.09109553 0.08041676 0.06756605 0.101748 0.07009308
2 0.09109553 0.08041676 0.06756605 0.101748 0.07009308
3 0.09109553 0.08041676 0.06756605 0.101748 0.07009308
4 0.09109553 0.08041676 0.06756605 0.101748 0.07009308
5 0.09109553 0.08041676 0.06756605 0.101748 0.07009308
6 0.09109553 0.08041676 0.06756605 0.101748 0.07009308
\end{Soutput}
\end{Schunk}
Supplying argument \code{prob} gives simulations that represent EVD quantiles. The above can be modified to give \code{nsim = 5} samples from the 100-year return level's sampling distribution for each row of \code{COprcp_plot} with
\begin{Schunk}
\begin{Sinput}
R> gev_rl_sim <- simulate(m_gev, nsim = 5, newdata = COprcp_plot, prob = 0.99)
R> head(gev_rl_sim)
\end{Sinput}
\begin{Soutput}
      [,1]     [,2]     [,3]     [,4]     [,5]
1 35.44633 39.64971 40.75644 49.70660 52.63366
2 36.08351 40.44809 41.55100 50.23721 53.64061
3 36.72078 41.24993 42.31399 50.78746 54.64288
4 37.35875 42.07773 43.11350 51.33915 55.66513
5 37.99928 42.91764 43.90430 51.90407 56.69209
6 38.62252 43.78138 44.72970 52.47980 57.72781
\end{Soutput}
\end{Schunk}
Suppose that a 95\% confidence interval for the 100-year return level for the third station, Boulder, in \code{COprcp_meta} is sought. This can be approximately achieved by estimating quantiles of the sampling distribution of the 100-year return level estimate. A 10,000-member sample can be drawn from this distribution with
\begin{Schunk}
\begin{Sinput}
R> gev_rl_boulder_sim <- simulate(m_gev, nsim = 1e4, newdata = COprcp_meta[3,], 
+    prob = 0.99)
\end{Sinput}
\end{Schunk}
and then its 2.5th and 97.5th empirical percentiles used to form an approximate 95\% confidence interval using
\begin{Schunk}
\begin{Sinput}
R> quantile(gev_rl_boulder_sim, c(0.025, 0.975))
\end{Sinput}
\begin{Soutput}
     2.5%     97.5% 
 97.61941 116.51000 
\end{Soutput}
\end{Schunk}
This could have been achieved with \fct{predict} using \code{se.fit = TRUE} if a symmetric sampling distribution was a fair assumption.

\subsubsection{Simulations of numerically-estimated return levels}

Approximate confidence intervals can also be obtained for numerically-estimated return levels. This is demonstrated for the example of \S\ref{FC:dy}, which uses Eq. \eqref{Fann_gpd}. First, parameters are simulated from the ALD and GPD models for each row in \code{rl_df}, introduced in \S\ref{FC:dy}, using
\begin{Schunk}
\begin{Sinput}
R> FC_sim_ald <- simulate(FC_ald, newdata = rl_df, nsim = 1e3, type = "response")
R> FC_sim_gpd <- simulate(FC_gpd, newdata = rl_df, nsim = 1e3, type = "response")
\end{Sinput}
\end{Schunk}
which gives 1000 samples. Then the 100-year return level is calculated for each sample using
\begin{Schunk}
\begin{Sinput}
R> rl_sim <- qev(0.99, FC_sim_ald[[1]], FC_sim_gpd[[1]], FC_sim_gpd[[2]], 
+    m = 365.25, theta = theta, family = "gpd", tau = 1 - zeta)
\end{Sinput}
\end{Schunk}
Again, the 2.5th and 97.5th percentiles estimated from the return level sample can be used to form an approximate 95\% confidence interval using
\begin{Schunk}
\begin{Sinput}
R> quantile(rl_sim, c(0.025, 0.975))
\end{Sinput}
\begin{Soutput}
    2.5%    97.5% 
38.63545 39.71674 
\end{Soutput}
\end{Schunk}
Note that 39.37$^\circ$C, the estimate obtained earlier from fitting separate GEV distributions to monthly maxima, falls well within this interval. Note also that uncertainty in the extremal index estimate, \code{theta} calculated in \S\ref{FC:dy}, is not propagated here.


\section{Summary and discussion} \label{sec:summary}

\begin{leftbar}
The \proglang{R} package \pkg{evgam} has been developed to allow the fitting of various EVDs with parameters of GAM form. Such forms are an intuitive and robust way of allowing parameters to vary with covariates. Examples in which parameters vary over space, through two-dimensional thin plate plates or the tensor product of two one-dimensional splines, and with time, specifically over the course of a year such that continuity is imposed from year to year, have been given. Examples also demonstrate fitting GEVs and GPDs, the Poisson-GPD model for extremes, and use of the ALD for threshold estimation through quantile regression. Various options for prediction and uncertainty estimation relevant to extreme value analyses have also been presented. Further functionality is planned for \pkg{evgam}.
\end{leftbar}


\section*{Computational details}


The results in this paper were obtained using
\proglang{R}~4.0.3 with the
\pkg{evgam}~0.1.4 package. \proglang{R} itself
and \pkg{evgam} are available from the Comprehensive
\proglang{R} Archive Network (CRAN) at
\url{https://CRAN.R-project.org/}.

\section*{Acknowledgments}

\begin{leftbar}
I thank an Editor, Reviewer and Yousra El Bachir for comments that have brought improvements to this article and \pkg{evgam} and Simon Brown, Steven Chan and Rob Shooter for highlighting bugs and/or functionality improvements that have improved \pkg{evgam}.
\end{leftbar}


\bibliography{/home/ben/mega/Mendeley/evgam.bib}

\begin{thebibliography}{38}
\newcommand{\enquote}[1]{``#1''}
\providecommand{\natexlab}[1]{#1}
\providecommand{\url}[1]{\texttt{#1}}
\providecommand{\urlprefix}{URL }
\expandafter\ifx\csname urlstyle\endcsname\relax
  \providecommand{\doi}[1]{doi:\discretionary{}{}{}#1}\else
  \providecommand{\doi}{doi:\discretionary{}{}{}\begingroup
  \urlstyle{rm}\Url}\fi
\providecommand{\eprint}[2][]{\url{#2}}

\bibitem[{Belzile \emph{et~al.}(2020)Belzile, Wadsworth, Northrop, Grimshaw,
  and Huser}]{mev}
Belzile L, Wadsworth JL, Northrop PJ, Grimshaw SD, Huser R (2020).
\newblock \emph{{mev: Multivariate Extreme Value Distributions}}.
\newblock \urlprefix\url{https://cran.r-project.org/package=mev}.

\bibitem[{Chavez-Demoulin and Davison(2005)}]{chavez2005}
Chavez-Demoulin V, Davison AC (2005).
\newblock \enquote{{Generalized additive modelling of sample extremes}.}
\newblock \emph{Journal of the Royal Statistical Society: Series C (Applied
  Statistics)}, \textbf{54}(1), 207--222.
\newblock \doi{10.1111/j.1467-9876.2005.00479.x}.
\newblock
  \urlprefix\url{https://rss.onlinelibrary.wiley.com/doi/abs/10.1111/j.1467-9876.2005.00479.x}.

\bibitem[{Coles(2001)}]{coles2001}
Coles SG (2001).
\newblock \emph{{An Introduction to Statistical Modeling of Extreme Values}}.
\newblock Springer-Verlag London.

\bibitem[{Cooley \emph{et~al.}(2007)Cooley, Nychka, and Naveau}]{cooley2007}
Cooley D, Nychka D, Naveau P (2007).
\newblock \enquote{{Bayesian spatial modeling of extreme precipitation return
  levels}.}
\newblock \emph{Journal of the American Statistical Association},
  \textbf{102}(479), 824--840.

\bibitem[{Davison and Ramesh(2000)}]{davison2000}
Davison AC, Ramesh NI (2000).
\newblock \enquote{{Local likelihood smoothing of sample extremes}.}
\newblock \emph{Journal of the Royal Statistical Society: Series B (Statistical
  Methodology)}, \textbf{62}(1), 191--208.
\newblock \doi{10.1111/1467-9868.00228}.
\newblock
  \urlprefix\url{https://rss.onlinelibrary.wiley.com/doi/abs/10.1111/1467-9868.00228}.

\bibitem[{Davison and Smith(1990)}]{davison1990}
Davison AC, Smith RL (1990).
\newblock \enquote{{Models for Exceedances Over High Thresholds}.}
\newblock \emph{Journal of the Royal Statistical Society: Series B
  (Methodological)}, \textbf{52}(3), 393--425.
\newblock ISSN 00359246.
\newblock \doi{10.1111/j.2517-6161.1990.tb01796.x}.
\newblock
  \urlprefix\url{http://doi.wiley.com/10.1111/j.2517-6161.1990.tb01796.x}.

\bibitem[{{De Boor}(1978)}]{deboor1978}
{De Boor} C (1978).
\newblock \emph{{A practical guide to splines}}.
\newblock Springer.

\bibitem[{Dutang and Jaunatre(2020)}]{dutang2020}
Dutang C, Jaunatre K (2020).
\newblock \enquote{{CRAN Task View: Extreme Value Analysis}.}
\newblock
  \urlprefix\url{https://cran.r-project.org/web/views/ExtremeValue.html}.

\bibitem[{Fasiolo \emph{et~al.}(2020)Fasiolo, Wood, Zaffran, Nedellec, and
  Goude}]{fasiolo2020}
Fasiolo M, Wood SN, Zaffran M, Nedellec R, Goude Y (2020).
\newblock \enquote{{Fast Calibrated Additive Quantile Regression}.}
\newblock \emph{Journal of the American Statistical Association},
  \textbf{0}(0), 1--11.
\newblock ISSN 0162-1459.
\newblock \doi{10.1080/01621459.2020.1725521}.
\newblock
  \urlprefix\url{https://www.tandfonline.com/doi/full/10.1080/01621459.2020.1725521}.

\bibitem[{Ferro and Segers(2003)}]{ferro2003}
Ferro CAT, Segers J (2003).
\newblock \enquote{{Inference for clusters of extreme values}.}
\newblock \emph{Journal of the Royal Statistical Society: Series B (Statistical
  Methodology)}, \textbf{65}(2), 545--556.
\newblock \doi{10.1111/1467-9868.00401}.

\bibitem[{Gilleland and Katz(2016)}]{gilleland2016}
Gilleland E, Katz RW (2016).
\newblock \enquote{{{\{}extRemes{\}} 2.0: An Extreme Value Analysis Package in
  {\{}R{\}}}.}
\newblock \emph{Journal of Statistical Software}, \textbf{72}(8), 1--39.
\newblock \doi{10.18637/jss.v072.i08}.

\bibitem[{Gilleland \emph{et~al.}(2013)Gilleland, Ribatet, and
  Stephenson}]{gilleland2013}
Gilleland E, Ribatet M, Stephenson AG (2013).
\newblock \enquote{{A software review for extreme value analysis}.}
\newblock \emph{Extremes}, \textbf{16}(1), 103--119.
\newblock ISSN 1386-1999.
\newblock \doi{10.1007/s10687-012-0155-0}.
\newblock \urlprefix\url{http://link.springer.com/10.1007/s10687-012-0155-0}.

\bibitem[{Green and Silverman(1994)}]{green1994}
Green PJ, Silverman BW (1994).
\newblock \emph{{Nonparametric regression and generalized linear models: a
  roughness penalty approach}}.
\newblock CRC Press.

\bibitem[{Hall and Tajvidi(2000)}]{hall2000}
Hall P, Tajvidi N (2000).
\newblock \enquote{{Nonparametric Analysis of Temporal Trend When Fitting
  Parametric Models to Extreme­Value Data}.}
\newblock \emph{Statist. Sci.}, \textbf{15}(2), 153--167.
\newblock \doi{10.1214/ss/1009212755}.
\newblock \urlprefix\url{https://doi.org/10.1214/ss/1009212755}.

\bibitem[{Heffernan and Stephenson(2016)}]{ismev}
Heffernan JE, Stephenson AG (2016).
\newblock \emph{{ismev: An Introduction to Statistical Modeling of Extreme
  Values}}.
\newblock \urlprefix\url{https://cran.r-project.org/package=ismev}.

\bibitem[{Koenker(2005)}]{koenker2005}
Koenker R (2005).
\newblock \emph{{Quantile Regression}}.
\newblock Econometric Society Monographs. Cambridge University Press.
\newblock \doi{10.1017/CBO9780511754098}.

\bibitem[{Koenker(2020)}]{quantreg}
Koenker R (2020).
\newblock \emph{{quantreg: Quantile Regression}}.
\newblock \urlprefix\url{https://cran.r-project.org/package=quantreg}.

\bibitem[{Northrop and Jonathan(2011)}]{northrop2011}
Northrop PJ, Jonathan P (2011).
\newblock \enquote{{Threshold modelling of spatially dependent non-stationary
  extremes with application to hurricane-induced wave heights}.}
\newblock \emph{Environmetrics}, \textbf{22}(7), 799--809.

\bibitem[{Oh \emph{et~al.}(2011)Oh, Lee, and Nychka}]{oh2011}
Oh HS, Lee TCM, Nychka DW (2011).
\newblock \enquote{{Fast Nonparametric Quantile Regression With Arbitrary
  Smoothing Methods}.}
\newblock \emph{Journal of Computational and Graphical Statistics},
  \textbf{20}(2), 510--526.
\newblock \doi{10.1198/jcgs.2010.10063}.

\bibitem[{Pauli and Coles(2001)}]{pauli2001}
Pauli F, Coles SG (2001).
\newblock \enquote{{Penalized likelihood inference in extreme value analyses}.}
\newblock \emph{Journal of Applied Statistics}, \textbf{28}(5), 547--560.
\newblock \doi{10.1080/02664760120047889}.
\newblock \urlprefix\url{https://doi.org/10.1080/02664760120047889}.

\bibitem[{Pfaff and McNeil(2018)}]{evir}
Pfaff B, McNeil A (2018).
\newblock \emph{{evir: Extreme Values in R}}.
\newblock \urlprefix\url{https://cran.r-project.org/package=evir}.

\bibitem[{Pickands(1971)}]{pickands1971}
Pickands J (1971).
\newblock \enquote{{The two-dimensional Poisson process and extremal
  processes}.}
\newblock \emph{Journal of applied Probability}, \textbf{8}(4), 745--756.

\bibitem[{{R Core Team}(2020)}]{R}
{R Core Team} (2020).
\newblock \emph{{R: A Language and Environment for Statistical Computing}}.
\newblock R Foundation for Statistical Computing, Vienna, Austria.
\newblock \urlprefix\url{https://www.r-project.org/}.

\bibitem[{Ramesh and Davison(2002)}]{ramesh2002}
Ramesh NI, Davison AC (2002).
\newblock \enquote{{Local models for exploratory analysis of hydrological
  extremes}.}
\newblock \emph{Journal of Hydrology}, \textbf{256}(1), 106--119.
\newblock ISSN 0022-1694.
\newblock \doi{https://doi.org/10.1016/S0022-1694(01)00522-4}.
\newblock
  \urlprefix\url{http://www.sciencedirect.com/science/article/pii/S0022169401005224}.

\bibitem[{Randell \emph{et~al.}(2016)Randell, Turnbull, Ewans, and
  Jonathan}]{randell2016}
Randell D, Turnbull K, Ewans K, Jonathan P (2016).
\newblock \enquote{{Bayesian inference for nonstationary marginal extremes}.}
\newblock \emph{Environmetrics}, \textbf{27}(7), 439--450.
\newblock ISSN 1099095X.
\newblock \doi{10.1002/env.2403}.

\bibitem[{Ribatet(2017)}]{SpatialExtremes}
Ribatet M (2017).
\newblock \emph{{SpatialExtremes: Modelling Spatial Extremes}}.
\newblock \urlprefix\url{https://cran.r-project.org/package=SpatialExtremes}.

\bibitem[{Rigby and Stasinopoulos(2005)}]{rigby2005}
Rigby RA, Stasinopoulos DM (2005).
\newblock \enquote{{Generalized additive models for location, scale and
  shape,(with discussion)}.}
\newblock \emph{Applied Statistics}, \textbf{54}, 507--554.

\bibitem[{Rue \emph{et~al.}(2009)Rue, Martino, and Chopin}]{rue2009}
Rue H, Martino S, Chopin N (2009).
\newblock \enquote{{Approximate Bayesian inference for latent Gaussian models
  by using integrated nested Laplace approximations}.}
\newblock \emph{Journal of the royal statistical society: Series b (statistical
  methodology)}, \textbf{71}(2), 319--392.

\bibitem[{Smith(1986)}]{smith1986}
Smith RL (1986).
\newblock \enquote{{Extreme value theory based on the r largest annual
  events}.}
\newblock \emph{Journal of Hydrology}, \textbf{86}(1), 27--43.
\newblock ISSN 0022-1694.
\newblock \doi{https://doi.org/10.1016/0022-1694(86)90004-1}.
\newblock
  \urlprefix\url{http://www.sciencedirect.com/science/article/pii/0022169486900041}.

\bibitem[{Smith(1989)}]{smith1989}
Smith RL (1989).
\newblock \enquote{{Extreme value analysis of environmental time series: an
  application to trend detection in ground-level ozone}.}
\newblock \emph{Statistical Science}, \textbf{4}(4), 367--377.

\bibitem[{Stephenson(2002)}]{evd}
Stephenson AG (2002).
\newblock \enquote{{evd: Extreme Value Distributions}.}
\newblock \emph{R News}, \textbf{2}(2), 0.
\newblock \urlprefix\url{http://cran.r-project.org/doc/Rnews/}.

\bibitem[{Wood(2006)}]{wood2006b}
Wood SN (2006).
\newblock \enquote{{Low-Rank Scale-Invariant Tensor Product Smooths for
  Generalized Additive Mixed Models}.}
\newblock \emph{Biometrics}, \textbf{62}(4), 1025--1036.
\newblock \doi{10.1111/j.1541-0420.2006.00574.x}.
\newblock
  \urlprefix\url{https://onlinelibrary.wiley.com/doi/abs/10.1111/j.1541-0420.2006.00574.x}.

\bibitem[{Wood(2011)}]{wood2011}
Wood SN (2011).
\newblock \enquote{{Fast stable restricted maximum likelihood and marginal
  likelihood estimation of semiparametric generalized linear models}.}
\newblock \emph{Journal of the Royal Statistical Society: Series B (Statistical
  Methodology)}, \textbf{73}(1), 3--36.

\bibitem[{Wood \emph{et~al.}(2016)Wood, Pya, and S{\"{a}}fken}]{wood2016}
Wood SN, Pya N, S{\"{a}}fken B (2016).
\newblock \enquote{{Smoothing parameter and model selection for general smooth
  models}.}
\newblock \emph{Journal of the American Statistical Association},
  \textbf{111}(516), 1548--1563.

\bibitem[{Yee and Stephenson(2007)}]{yee2007}
Yee TW, Stephenson AG (2007).
\newblock \enquote{{Vector generalized linear and additive extreme value
  models}.}
\newblock \emph{Extremes}, \textbf{10}(1-2), 1--19.
\newblock ISSN 13861999.
\newblock \doi{10.1007/s10687-007-0032-4}.

\bibitem[{Yee and Wild(1996)}]{yee1996}
Yee TW, Wild CJ (1996).
\newblock \enquote{{Vector Generalized Additive Models}.}
\newblock \emph{Journal of the Royal Statistical Society: Series B
  (Methodological)}, \textbf{58}(3), 481--493.
\newblock \doi{10.1111/j.2517-6161.1996.tb02095.x}.
\newblock
  \urlprefix\url{https://rss.onlinelibrary.wiley.com/doi/abs/10.1111/j.2517-6161.1996.tb02095.x}.

\bibitem[{Youngman(2019)}]{y2019}
Youngman BD (2019).
\newblock \enquote{{Generalized Additive Models for Exceedances of High
  Thresholds With an Application to Return Level Estimation for U.S. Wind
  Gusts}.}
\newblock \emph{Journal of the American Statistical Association},
  \textbf{114}(528), 1865--1879.
\newblock ISSN 0162-1459.
\newblock \doi{10.1080/01621459.2018.1529596}.
\newblock
  \urlprefix\url{https://www.tandfonline.com/doi/full/10.1080/01621459.2018.1529596}.

\bibitem[{Yu and Moyeed(2001)}]{yu2001}
Yu K, Moyeed RA (2001).
\newblock \enquote{{Bayesian quantile regression}.}
\newblock \emph{Statistics {\&} Probability Letters}, \textbf{54}(4), 437--447.
\newblock ISSN 0167-7152.
\newblock \doi{10.1016/S0167-7152(01)00124-9}.

\end{thebibliography}

\end{document}